# Transmission resonance in a composite plasmonic structure

Xiao-gang Yin, Cheng-ping Huang, Qian-jin Wang, Chao Zhang,

and Yong-yuan Zhu*

*National Laboratory of Solid State Microstructures, Nanjing University*

*Nanjing 210093, P.R. China*

## Abstract

The design, fabrication, and optical properties of a composite plasmonic structure, a two-dimentional array of split-ring resonators inserted into periodic square holes of a metal film, have been reported. A new type of transmission resonance, which makes a significant difference from the conventional peaks, has been suggested both theoretically and experimentally. To understand this effect, a mechanism of ring-resonance induced dipole emission is proposed.

Email: yyzhu@nju.edu.cn

The interaction of light with two-dimensional (2D) arrays of periodic holes or particles has sparked great interest [1]. These holes or particles are usually of the sub-wavelength scale and possess anomalous transmission properties. It is well known that a metal film with subwavelength holes has the ability to enhance the transmission of light at certain wavelengths [2], where the zero-order transmission can be larger than unity when normalized to the area of the holes. Generally, it is admitted that the surface-plasmon polariton (SPP) mode originating from coupling of light to collective oscillation of electrons will play a crucial role [3, 4]. Nonetheless, theoretical calculations show that it is the diffracted modes created by the periodicity of hole lattice (the parallel wavevector of diffracted modes equals the parallel component of incident wavevector plus a reciprocal lattice vector) that accounts for the unusual effect [5, 6]. In spite of the debate on the mechanism, this effect has a wide range of potential applications in such as nanolithography, quantum optics etc. [7].

On the other hand, the metallic particles can support the localized plasmon resonance with the resonance frequency determined by the particle shape and size. Especially, when the particle is made to be a split-ring resonator (SRR), a ring resonance will be established under the action of an incident light [8-13]. Note that the multiple ring resonances can be understood as plasmonic resonances of the entire SRR structure with different orders [8, 9]. The fundamental plasmonic resonance, corresponding to a circular electric current in the SRR, is dominant and usually known as an LC resonance. Such a resonance can be excited by the light magnetic field [10, 11], and in that case metamaterials with negative permeability has been demonstrated. More recently, with the use of the electric field of a normally incident light, the excitation of ring resonance of SRRs has also been realized [12, 13]. It was found that a 2D array of SRRs can suppress the transmission of light, which occurs at the resonance frequencies of the single SRRs.

In this paper, the design, fabrication, and optical properties of a composite plasmonic structure, a 2D array of SRRs inserted into the periodic square holes of a metal film, have been reported. With the new structure, we have realized an extraordinary optical transmission effect. Compared with previous work on light transmission [2-6, 12, 13], our results make two significant differences. Firstly, in addition to the conventional transmission peaks, we have also observed a new type of transmission resonance at a longer wavelength. The two types of resonances rely on the structural parameters in different way. Secondly, corresponding to the fundamental ring resonance, a switching of light transmission from suppression in the pure SRRs to enhancement in the present structure has been suggested. A mechanism of ring-resonance induced dipole emission accounts for the effect.

The composite plasmonic structure under investigation is shown schematically in Fig. 1(a), and the corresponding structural parameters of the unit cell are shown in Fig. 1(b). Here, a metal film with the thickness $t$ is deposited onto a glass substrate. Square holes with the side length $a$ are cut into the film in a square array of lattice constant $p$. And, the U-shaped SRRs with the bottom length $b$, arm length $d$, and gap width $w$, are inserted into each square hole. An x- or y-polarized light is incident normally upon the metal film and transmits from air to the substrate. Here x- or y-polarization means that the electric field of incident light is along the x- or y-axis, as shown in Fig. 1(b).

To reveal the optical properties of the plasmonic structure, numerical simulations have been performed with the finite-difference time-domain (FDTD) method [14]. In the simulations, the permittivity of air and glass substrate is set as 1 and 2.25, respectively, and the metal (silver) is modeled by the Drude dispersion with a plasma frequency of $\omega_p = 1.37 \times 10^{16}$ rad/s and a collision frequency of $\gamma = 2 \times 10^{14}$ rad/s. The latter is larger than that of the bulk metal, due to additional scattering from the nanoscale metal surface [15, 16].

Without loss of generality, a composite structure with the film thickness t=70nm, lattice period p=600 nm, and hole side a=300 nm has been studied, where the sizes of SRR are set as b=170 nm, d=100 nm, and w=50 nm. To start, Fig. 2(a) has presented the calculated zero-order transmission spectra of a pure hole array (without SRRs, the black line) and that of a pure SRR array (the red and green lines are for x- and y-polarization, respectively). The black line in Fig. 2(a) exhibits a set of transmission peaks as observed previously [2, 17], which can be attributed to the so-called SPP resonance on the metal-dielectric interface. Meanwhile, the transmission dips in the red and green lines are associated with the ring resonances of SRR [8], where the fundamental resonance is around 1220nm.

However, when the SRRs are inserted into the square holes, new features appear in the transmission spectrum. Figure 2(b) shows the calculated transmission spectra of the composite structure for both x (the red line) and y-polarization (the green line). One can see that the spectra of the composite structure are characterized by the transmission maxima but modified compared with that of a pure hole array. Especially, for the x-polarization, besides the transmission peaks corresponding to the wavelength of SPP resonance, a new transmission peak appears on the longer wavelength side of the spectrum (around the telecom wavelength 1500nm, as marked by an arrow). Apparently, the position of this new peak is far beyond the wavelength of Ag-glass (1, 0) SPP resonance (at 1000 nm). In addition, when the incident light is y-polarized, the spectrum does not show any essential changes. Nonetheless, the resonance peaks at 800 and 1000nm are red-shifted, respectively, to around 900 and 1200nm (due to a coupling between SPP mode and high-order plasmonic mode of SRR). It should be pointed out that the spectra of the composite structure are different from the response of complementary SRR [18-20]. This can be found by checking the spectra in Fig. 2(a) and Fig. 2(b), where one-to-one correspondence between the minima of pure SRRs and the maxima of composite structure is not present.

The theoretical prediction of the composite structure has been demonstrated by the experiments. We coated the polished K9-glass with a 70 nm thick silver film (5 nm thick Ge film was used as an adhesion layer) and fabricated in the film the designed structure using the focused-ion-beam (FIB) system (Strata FIB 201, FEI Company, 30 KeV Ga ions, 4pA beam current). The typical FIB image of a part of the fabricated sample (the array consists of 66×66 units) is shown in Fig. 1(c). In the measurement, a homebuilt optical setup is utilized to analyze the transmission properties of the sample. After a slight focalization, the white x- or y-polarized light from a 50 W halogen lamp (wavelength ranging from 400 nm to 1800 nm) impinges normally onto the sample. The zero-order transmission intensities are collected by an optical spectrum analyzer (ANDO AQ-6315A) (the transmission of the bare glass substrate has been used as a reference). The measured transmission spectra of the composite structure for the two incident polarizations are shown in Fig. 2(c). It can be seen that the experimental results agree well with the theoretical calculations, concerning the spectra shape, the peak position and peak width.

We conclude that in the x-polarization the shorter-wavelength peaks of the composite structure are mainly dominated by and can be attributed to the SPP resonance, which are slightly modified in the presence of SRRs. But, what is the origin of the new transmission peak? To answer the question, we have plotted with the FDTD method the current density distribution (Fig. 3(a)) and the magnetic field pattern (Fig. 3(b)) on the plane z=-35nm, for the new transmission peak marked in Fig. 2. Figure 3(a) clearly shows that a distinct current loop appears in the SRR. This indicates that the fundamental ring resonance of the SRR is unambiguously excited at the wavelength of the new peak [9]. This point can also be reinforced by the magnetic field pattern showed in Fig. 3(b), where the magnetic field generated by the circular electric current is strongly localized in the SRR air gap. In this case, the light is normally incident on the sample with the electric field along the bottom of the SRR. It is the electric field of the incident light that excites the fundamental ring resonance [9]. Therefore, the appearance of the new peak is related to the fundamental ring resonance of the SRRs inserted into the square holes.

For a pure SRR array, the ring resonance causes the incident light to be absorbed and scattered, resulting in a dip as observed in the transmission spectrum [12, 13]. But, why is here a transmission peak rather than a dip for the ring resonance? Let us see the figure 3(a), again. From figure 3(a), one can see that besides the circular electric current in the SRR, the strong localized currents also appear in the metallic film around the square hole. Furthermore, the current in the SRR and that around the upper half of the hole form one current loop, whereas the current in the bottom of the SRR and that around the lower half of the hole form another loop. This clearly indicates that the ring resonance of the SRR interacts strongly with the electrons in the surrounding metallic film. As a consequence, the positive and negative electric charges are induced and localized at the opposite sides of the air gaps, forming the oscillating electric dipoles (here air gaps refer to the left and right gaps between the hole walls and SRR). That is to say, the electric dipoles are induced by the ring resonance of the SRR. These oscillating dipoles will emit the electromagnetic waves into the far field, leading to the formation of new peak in the transmission spectrum. It is worthy of noticing that, due to the interaction between SRR and the surrounding metallic film, the ring resonance frequency of the composite structure has been modified compared with that of a pure SRR array (see Fig. 2). This agrees with the further experimental and theoretical findings that the peak of ring resonance will vary with the hole side length when the sizes of SRR are fixed (not shown here). The smaller the hole size is, the stronger the interaction and the larger the resonance wavelength. It is also significant that the induced electric dipoles can emit the radiation back to the incident side, which interferes destructively with the light reflected from the metal surface. Hence, the reflected light will be suppressed, giving rise to a reflection minimum at the fundamental ring resonance. This point has been confirmed by the numerical calculations.

To get more insight into the phenomenon, we have studied the dependence of zero-order transmission on the geometrical parameters of the composite structure (for x-polarization). Figure 4(a) and 4(b) presents, respectively, the calculated and measured transmission spectra with different arm length, *d*, of the SRR. Here the arm length is increased from 80nm to 100nm and 120nm, and the remaining parameters are fixed. Note that the theory and experiment agrees well with each other. As *d* increases, the positions of SPP peaks do not vary significantly. Nonetheless, the peak of fundamental ring resonance shifts apparently to the longer wavelength (from around 1470nm to 1560nm and 1660nm). This is in accordance with the character observed in a pure SRR array [13] and can be understood using an LC circuit model (the ring resonance frequency of a single SRR can be expressed as $\omega = 1/\sqrt{LC}$, where $L$ and $C$ is the inductance and capacitance of SRR, respectively. Since $L$ is proportional to the total length of SRR $l = b + 2d$ and C is independent of the arm length $d$, the resonance wavelength $\lambda \propto \sqrt{b+2d}$ will then increase with the arm length). Figure 4(c) and 4(d) shows, respectively, the calculated and measured transmission spectra for three lattice period $p$=500, 600 and 700nm. Again, good agreement between theory and experiment is resulted. As $p$ increases, the positions of SPP peaks shift to the longer wavelength as expected, but the peak of ring resonance does not. This suggests that the ring resonance is mainly a character of the individual unit cells rather than that of the periodic array, thus corresponding to a localized mode. Similar simulations also show that the spectral position of ring resonance is not sensitive to the incident angle, which means the ring resonance has a flat dispersion. Additionally, we have studied the dependence of ring resonance on the film thickness. An increase of the film thickness corresponds to a significant blue-shift of the ring resonance peak. A similar effect has been suggested in a pure SRR array and explained in terms of an equivalent cut-wire model [21].

Our results suggest a dual-channel for the light transmission through the proposed composite structure. When the light is incident normally onto the structure, the optical field captured inside the periodic holes can be enhanced by the SPP resonance, leading to the conventional transmission peaks. On the longer wavelength side where the SPP resonance cannot be excited, however, the incident light may couple to the fundamental ring resonance of the SRRs. The ring resonance drives the free electrons around the individual holes, giving rise to the oscillating electric dipoles. These electric dipoles will emit the radiation into the far field, resulting in a new type of transmission peak located at the longer wavelength. Since both extended SPP resonance and localized ring resonance can be supported, the structure proposed represents another example of the localized/delocalized system [22]. The availability of strong electric field (due to SPP resonance) and strong magnetic field (due to ring resonance), as well as extraordinary optical transmission, may find unique applications in plasmonics. The effect reported here may also be extended to low-frequency regions, such as THz or microwave band.

This work was supported by the State Key Program for Basic Research of China (Grant Nos. 2004CB619003 and 2006CB921804), by the National Natural Science Foundation of China (Grant No. 10523001, 10804051, and 10874079).

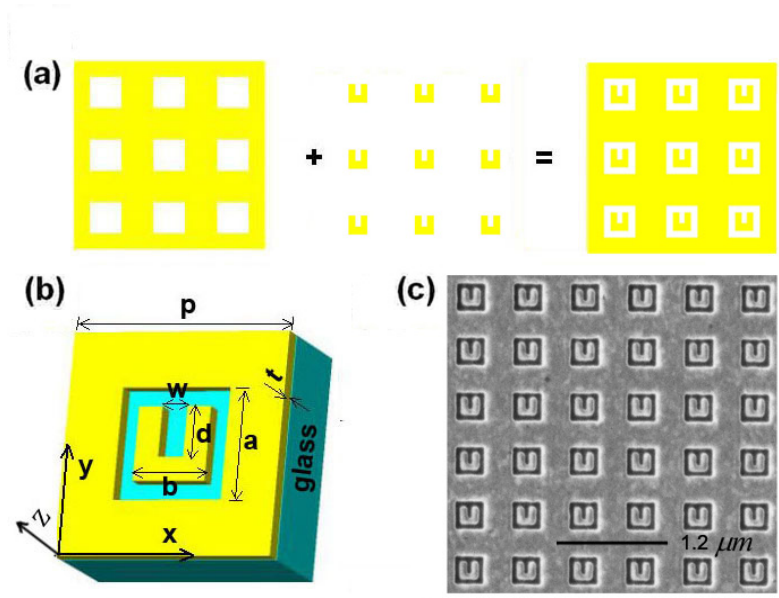

FIG.1. (Color online) (a) Scheme of the composite metamaterial design, (b) The unit cell with the corresponding structural parameters, and (c) The focused-ion-beam (FIB) image of a typical array of the composite structure (the lattice period is p=600 nm). All samples have a glass substrate and a metal thickness of t=70 nm.

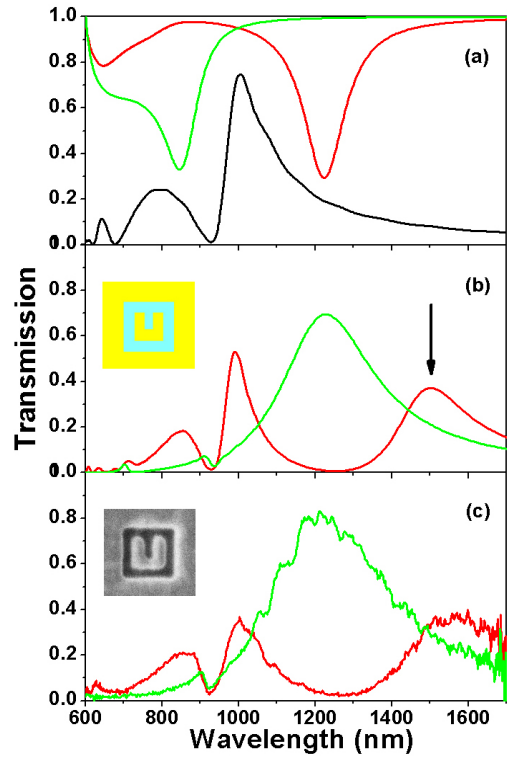

FIG.2. (Color online) Zero-order transmission spectra of the plasmonic structures: (a) Calculated results of a pure hole array (the black line) and a pure SRR array (the red and green lines are for x- and y-polarization, respectively). (b) Calculated and (c) measured spectra of the composite structure, for x-polarization (the red line) and y-polarization (the green line), respectively. Here, the lattice period is p=600 nm, the hole side length is a=300 nm, and the structural parameters of SRR are b=170 nm, w=50 nm, and d=100 nm.

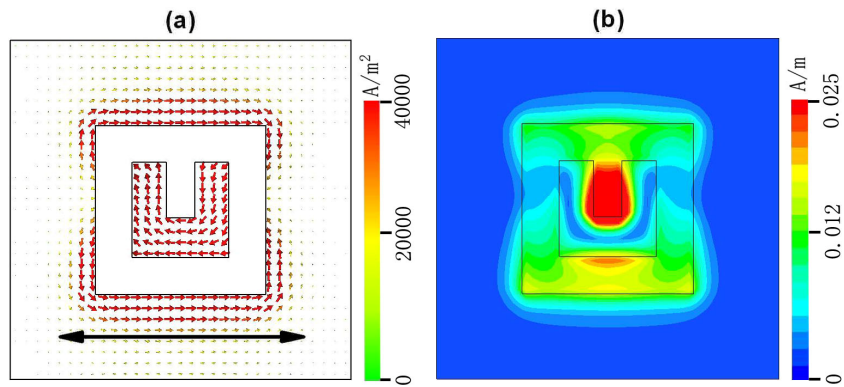

FIG.3. (Color online) Simulated current density (arrows) distribution (a) and the magnetic field pattern (b) on the x-y plane (z=-35nm), for the new transmission peak (marked by an arrow in Fig.2 (b)). The bold solid arrow specifies the incident polarization (x-polarization). The image covers a single period, and the black lines mark the profiles of the square holes and SRRs.

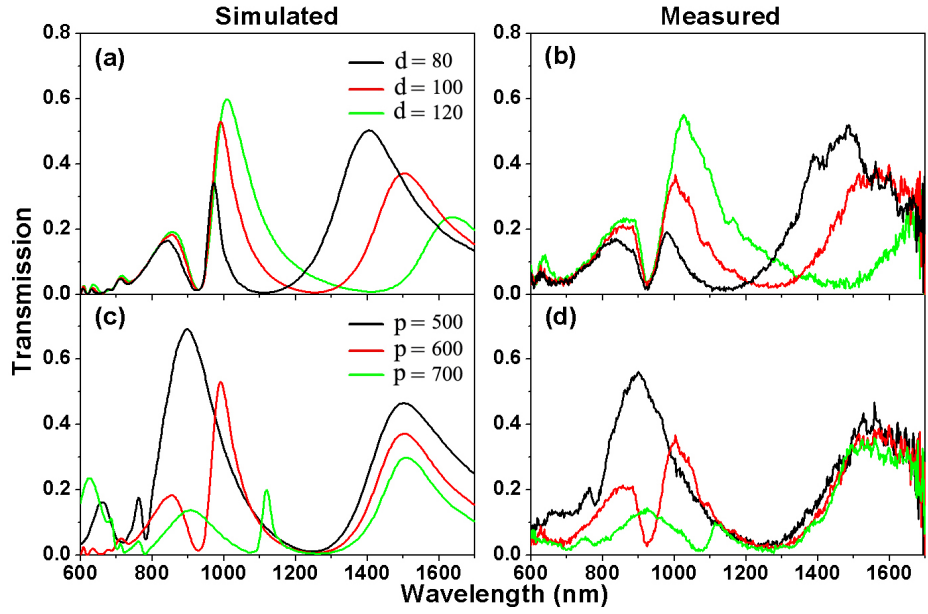

FIG.4. (Color online) Calculated (left column) and measured (right column) zero-order transmission spectra, for x-polarization. In the first row, the arm length, d, of the SRRs is changed from 80 to 100 and 120 nm, and the remaining parameters are fixed as p=600 nm, a=300nm, b=170 nm, and w=50 nm; In the second row, the lattice period p is increased from 500 to 600 and 700 nm, with the fixed parameters a=300nm, b=170 nm, w=50 nm, and d=100nm.